\documentclass[9pt,preprint2,natbib209]{aastex61}

\usepackage{graphicx}
\usepackage{dcolumn}
\usepackage{bm}
\usepackage{subfigure}

\def\apjs{Astrophys. J.  Supp. Seri.}

\def\apjl{Astrophys. J. Lett.}

\def\aap{Astro. \& Astrop.}

\begin{document}

\title{A Study of Wolf-Rayet stars Formed via Chemically Homogeneous Evolution}

\correspondingauthor{Zhanojun Wang, Guoliang L\"{u}}\email{guolianglv@xao.ac.ac, xjdxwzj@sohu.com}

\author{Zhe Cui}\email{zhecui01@163.com}
\affil{School of Physical Science and Technology,
Xinjiang University, Urumqi, 830046, China}

\author{Zhaojun Wang}
\affil{School of Physical Science and Technology,
Xinjiang University, Urumqi, 830046, China}

\author{Chunhua Zhu}
\affil{School of Physical Science and Technology,
Xinjiang University, Urumqi, 830046, China}

\author{Guoliang L\"{u}}
\affil{School of Physical Science and Technology,
Xinjiang University, Urumqi, 830046, China}

\author{Hailiang Chen}
\affil{Yunnan Observatories, the Chinese Academy of Sciences, Kunming 650011, China}

\author{Zhanwen Han}
\affil{Yunnan Observatories, the Chinese Academy of Sciences, Kunming 650011, China}

\begin{abstract}
Using the stellar evolution code---Modules for Experiments in Stellar Astrophysics (MESA), we investigate the evolution of
massive stars with different rotational velocities and metallicities towards Wolf-Rayet stars.
In our simulations, the initial rotating velocities are taken as 0, 250, 500 and 650 km s$^{-1}$,
and the metallicities equal to 0.02, 0.014, 0.008, 0.006, 0.004 and 0.002.
We show our rapid rotation models in the HR diagram compared with the observations.
We find that the rotational mixing is less efficient at high metallicity,
and these stars become Wolf-Rayet (WR) stars when the helium in their center
is ignited. However, rapid rotating massive stars at low metallicity can easily evolve into WR stars
due to the rotation resulted in chemically homogeneous evolution. This can explain the origin of single WR stars
in galaxy at low metallicity. In our models, the observed SMC WR stars are consistent with the single-star evolution models. However at higher metallicities our single-star evolution models can only explain the luminous, hydrogen-rich WN stars and O stars (which are classified as WR stars previously).

\end{abstract}

\keywords{stars: evolution---stars: rotation---stars: Wolf-Rayet}
\section{Introduction}

  Wolf-Rayet ($\rm WR$) stars are, typically, helium-burning stars who have lost a substantial
  part of their hydrogen envelope via stellar wind or mass transfer through Roche-lobe overflow
  in close-binary systems, and they are fusing helium or heavier elements
  in the core\citep{Chiosi1986, Maeder1994}. Usually, they are
  hot $(\rm \log(T_{eff}/K)\geq 4)$ and luminous stars $(\rm \log(L/L\odot)>5.0)$.
  Spectroscopically, distinguished from normal stars, $\rm WR$ stars are objects
  with strong, broad emission lines \citep{Beals1940},
  in which the broadening of lines is caused
  by the large expansion velocity in the expanding stellar wind and
  their emission characteristic is mainly because of the powerful deviations
  from local thermodynamical equilibrium \citep{Todt2015}.
  Based on the relative intensity of the spectrum,
  $\rm WR$ stars can be cursorily classified into three subtypes: $\rm WN$, $\rm WC$ and $\rm WO$,
  depending on whether the spectrum was dominated by lines of nitrogen, carbon or oxygen, respectively.
  $\rm WR$ stars with both $\rm WN$ and $\rm WC$ characteristic are classified as $\rm WN/WO$ stars.
  There are about 642 $\rm WR$ stars in our Galaxy, including 357 $\rm WN$ stars,
  273 $\rm WC$ stars, 8 $\rm WN/WC$ stars and 4 $\rm WO$ stars \citep{Crowther2015}.
  As one of the closest galaxies to Galaxy, the Large Magellanic Cloud
  ($\rm LMC$ ) allows a detailed spectroscopy of its brighter stars \citep{Hainich2014}.
  The forth catalogue of population I $\rm WR$ stars provides about 134 $\rm WR$ stars in
  LMC \citep{Breysacher1999}, and recently 13 more $\rm WR$ stars in the
  LMC were discovered by \cite{Massey2015}. There are 12 WR stars currently known
  in the Small Magellanic Cloud ($\rm SMC$) \citep{Massey2014}, and 5 out of the 12 stars are in confirmed binary
or multiple systems based on their RV curves ($\rm FMG$) \citep{Shenar2016}.

  $\rm WR$ stars are very important objects. They are
  dominant sources of energy and nuclear synthesis
  products such as helium (He), carbon (C), oxygen (O) and other $\alpha$ elements to their surroundings thanks
  to their strong stellar winds \citep{Eldridge2006}.
  They are also the progenitors of type Ib/c supernovae (SNe) due to their lack of H
  and long gamma-ray bursts (LGRBs)\citep{Galama1998,Ensman1988,Hjorth2003,Woosley2006}.
  And their evolution towards core collapse is critically determined by mass loss \citep{Yoon2017}.

  After years of observational and theoretical efforts,
  basic concepts about massive stars
  have been established and provide a good guide to their observed properties
  from the aspects of single star and binary system.
  One of the most known picture of stellar evolution is 'Conti scenario' \citep{Conti1975}.
  He first proposed that a massive O stars may lose a significant
  amount of mass of envelope through stellar wind and reveal the core
  H-burning or later He-burning products (if
  there is sufficient additional mass loss) at
  its surface, this evolutionary stages are spectroscopically
  identified with $\rm WN$ and $\rm WC$ types \citep{Crowther2007}.
  Nowadays, it is argued that this
  process is occasionally aided by Roche-lobe overflow
  in close binaries and/or episodic mass loss during
  the LBV stage \citep{Smith2006,Massey2015}.

  However, the true evolution conditions of $\rm WR$ stars are still not completely understood
  owing to some indeterminate issues, such as the interior mixing processes and stellar mass-loss processes
  which perplex the massive stellar evolutionary models all the time \citep{Chiosi1986,Hamann2006b}.
  The strong mass-loss rate
  ($\dot{M}$) of the order of $10^{-5} M_{\odot}{\rm yr}^{-1}$ is an important feature of massive stars \citep{vanderHucht2001}.
  It is closely related to the stellar luminosity and effective temperature
  with the relation that the higher the luminosity is,
  the more violent the mass-loss rate behavior \citep{Jaeger1988}.
  The prescriptions of $\rm WR$ mass-loss rates adopted
  in stellar evolutionary models before are very large \citep{Maeder1987,Langer1989,Hamann1995}.
  Although \cite{Meynet1994} obtained a good agreement with the observed WR populations for the
  metallicities ranging from that of the SMC to twice the solar metallicity at high mass loss rates.
  The consideration of wind clumping in later empirical estimates resulted in much lower values \citep{Hamann1998,Hamann2006a,Crowther2007,Sander2012}.
  One of the most striking results among these later studies is
  the prescription that mass-loss rate
  should be reduced by a factor of 2 to 3 than that applied before
  because $\rm WR$ winds are optically thick and inhomogeneous \citep{Nugis2000, Hamann1999}.

  Moreover, observations and theoretical researches indicate that
  stellar wind strongly depends on metallicity and
  behave as a power-low: $\dot{M}\propto Z^{m}$ in which $m$ is the index ranging from
  1/2 to 0.94 \citep{Garmany1985, Prinja1987, Castor1975, Abbott1982, Pauldrach1986, Vink2000, Vink2001}.
  Therefore, theoretically, mass-loss rates for $\rm WR$ stars at low metallicities
  would perform much lower than those at high metallicities.
  As a consequence, stellar wind will be too weak to strip off their
  H-rich envelope to evolve into $\rm WR$ phase. Some works suggest that
  $\rm WR$ stars at low metallicities could be formed by the means of
   mass transfer through Roche Lobe Overflow in close binary systems \citep{Bartzakos2001,Maeder1982,Vanbeveren1998}.

   Nevertheless, this conjecture was pushed down by the works
   of \cite{Foellmi2003a,Foellmi2003b} and \cite{Foellmi2004} which indicate
   that even at low metallicity, such as the SMC and LMC,
   a large fraction of the $\rm WR$ stars may originate via the
   single star scenario, similar to that in the Milky Way.
   Therefore another process must be at work.

    A practicable channel to form
    WR stars from single stars without invoking mass
    loss is supplied by the scenario of chemically
    homogeneous evolution \citep{Maeder1987,Langer1992,Yoon2005,Schootemeijer2018},
    which means stars evolving
    with a nearly uniform chemical composition from the centre to the surface.
    Homogeneously evolution can be triggered by various mechanisms \citep{Georgy2015}:
    a)internal mixing inside the stars induced by convective movements of
    materials in the convective regions \citep{Maeder1980, Yusof2013};
    b)mixing progress in the radiative regions, such as rotational mixing \citep{Zahn1992, Maeder1987}.
    The proposal of this scenario is attribute to the less role of mass loss in producing $\rm WR$ at low metallicity,
    and the observational evidence of the large scale structures harboured in certain $\rm WR$ stars
    also indicated that a rotating velocity may be existed \citep{StLouis2007, Crowther2007}.

    Besides, as one of the quantitative measurements of nuclear burning times
    and the importance of mass loss during various stages of the stars¡¯ lifetimes \citep{Eldridge2008},
    the reproduce of WC/WN number ratio of WR stars is
    momentous but difficult owing to our uncertainty of their surface temperatures and lumiosities.
    While the rotating models could reproduce them to some degree.
    \cite{Meynet2003} discussed the effects of rotation on $\rm WR$ stars at solar metallicity and found that the
    theoretical predictions of the number ratios of $\rm WR$ stars
    matched the observations well when the effects of
    rotation are accounted for. By contrast, the standard non-rotating models did not
    agree with these observed rates. The studies of massive single-star evolution
    considering both mass loss and rotation also
    show a well reproduction of the observed variation of type Ib/Ic SNe with
    respect to type II SNe fractions with metallicities\citep{Meynet2005}.
    Evolutionary models obtained from the new
    Geneva Population Synthesis code which take rotational
    mixing into account slighting the discrepancy between the synthetic and
    observed population than the older tracks without rotation \citep{Hamann2006a}.
    In addition, the works by mixing single and binary star populations
    got a better agreement between the observational
    values and their predicted values \citep{Vanbeveren2007,Eldridge2008}.

    It is clear that stellar rotation plays an essential role in massive star evolution,
    influencing the output such as stellar lifetimes, evolutionary tracks,
    surface abundances, pre-supernova status and even the deaths and contributions to
    interstellar medium \citep{Meynet2005, Maeder2010}.
    High rotational velocity of stars can be accompanied by their births or acquired
    through the accelerating mechanism induced by tidal forces, material accretion or merging of
    stars \citep{Petrovic2005a,Petrovic2005b,deMink2009,deMink2013,Tylenda2011,Dervisoglu2010, Song2016}.
    It is needed to provide the large number of core angular momentum
    required for explosion to become LGRBs as the death of the most
    massive stars especially at low metallicity \citep{Martins2013}.
   \cite{Georgy2012} suggests that about half of the
   observed WR stars and at least half of the Type Ib/Ic SNe may be
   produced through the single-star evolution channel predicted by
   their rotating stellar models at Z=0.014.

   Recently thanks to work by the Potsdam WR group we now have a much better
   understanding of where the WR stars in the HR diagram. Their comparison of
   tracks to the WR stars indicated that while models could reproduce the WC/WN
   ratio they could not reproduce locations in the HR diagram (e.g. \cite{Sander2012}).
   Lately \cite{Eldridge2017} have shown that with binary models they can reproduce
   the observed HR diagram locations as well as the WC/WN ratio at
   different metallcities. Also \cite{Shenar2016} have shown that the WR stars
   in the SMC can be reproduced by standard single-star models as well as
   binary models while binary models are required at higher metallicities to explain the luminosity.

   Considering above, in this work,
   we study how rotation lead to the quasi-chemically homogeneous evolution
   therefore modifies the evolution of a given initial mass
   star towards the $\rm WR$ phase in single-star evolutionary scenario.
   And we test our rapid rotating models against the observed WR locations in the HR diagram.

  In Section 2 a brief summary of the physics adopted in our models is given,
  in Section 3 we present the effects of rotation on stellar evolution and
  show the HR diagram of most known WR stars in MW, LMC and SMC,
  And a synthesis of the main results is presented in Section 4.
\section{Model}

 In this paper, we employ the open-source stellar evolution
 code $\rm MESA$ (version 8848, \cite{Paxton2011, Paxton2013, Paxton2015})
 to simulate the structure and evolution of rotating massive stars.
\cite{Brott2011} had produced several
 grids of evolutionary models for rotating massive stars.
Using similar parameters with those in \cite{Brott2011},
\cite{Zhu2017} investigated the effects of the core-collapse supernova ejecta on rotating massive
star.
Following \cite{Brott2011} and \cite{Zhu2017},
the Ledoux criterion
is used for convection, mixing-length parameter ($\alpha_{\rm LMT}$)
and an efficiency parameter ($\alpha_{\rm SEM}$) for
semi-convection are taken as 1.5 and 1.0, respectively.

The mass-loss rates we use in our code are the same with \cite{Brott2011}.
For stars hotter than about 25 kK with surface hydrogen mass fraction of $X_{S}>0.7$, we use the wind recipe of \cite{Vink2001}.
\cite{Vink2001} gave the formulae of mass-loss rate for massive stars.
 If the massive star is rapidly rotating, the mass-loss rate
 would be enhanced, which was given by \cite{Langer1998}
 \begin{equation}
\dot{M}=(\frac{1}{1-\Omega/\Omega_{\rm crit}})^\beta \dot{M}_{v_{\rm rot}=0},
\label{eq:ml}
\end{equation}
where $\Omega$ and $\Omega_{\rm crit}$ are the angular velocity and the critical angular velocity, respectively,
and $\beta=0.43$ \citep{Langer1998}.
For hydrogen-poor hot stars with $X_{S}<0.4$, we use the WR mass loss recipe
from \cite{Hamann1995}, reduced by a factor of ten.
The mass-loss rate results from a linear interpolation between \cite{Vink2001}
and \cite{Hamann1995} for $0.4<X_{S}<0.7$.
We use the highest of the values given from the prescriptions of \cite{Vink2001}
and \cite{Nieuwenhuijzen1990} when the stars cooler than the critical temperature
for the bi-stability jump ($\sim 25$k K).
When stars evolve into RSG phase (the central hydrogen is exhausted and the
effective is lower than 10k K), we use the mass-loss rate given by \cite{Nieuwenhuijzen1990},
which does not depend on metallicity.
Simultaneously, rotational mixing induces various instability,
such as dynamical shear instability, Solberg-Hi{\o}land instability,
secular shear instability, Eddington-Sweet circulation, and the
Goldreich-Schubert-Fricke instability \citep{Spiegel1970,Zahn1974,Zahn1975,Wasiutynski1946,Goldreich1967,Fricke1968,Endal1978,Pinsonneault1989,Heger2000}.
Considering these instabilities provides an alternative procedure to
restrict the class of angular-velocity distributions considered which is needed for the calculation of a static model \citep{Spiegel1970}.
Following \cite{Brott2011} and \cite{Zhu2017}, the ratio of the turbulent viscosity to the diffusion coefficient ($f_{\rm c}$)
and the ratio of sensitivity to chemical gradients ($f_{\rm \mu}$) are taken as 0.0228 and 0.1, respectively \citep{Heger2000,Yoon2006}.

Mass-loss rate and these instabilities are affected by the metallicity ($Z$) \citep{Heger2000}.
Considering the relevant metallicities used by most studies are Z=0.014 to 0.020 for the MW,
for the LMC Z=0.06 to 0.008 and Z=0.002 to 0.004 for the SMC,
we take the two critical values for each galaxies respectively, that is to say,
the initial abundance of hydrogen (X), helium (Y) and metal (Z)
 used in our models are : Z=0.02 and 0.014 for the MW, Z=0.008 and 0.006 for
 the LMC and Z=0.004 and 0.002 for the SMC,
 the corresponding initial helium mass fractions
 Y are given by the relation $Y = Y_{P} + \triangle Y/\triangle Z\cdot Z$,
 where $Y_{P} = 0.23$ and $\triangle Y/\triangle Z = 2.25$ are the the
 primordial helium abundance and slope of the helium-to-metal enrichment
 law respectively \citep{Meynet2005,Maeder2001}, X=1-Y-Z.
 Meanwhile,
 we also calculate three groups of models by referring to the work
 of \cite{Brott2011} : $\rm X = 0.7274$, $Y = 0.2638$, $Z = 0.0088$,
 $\rm X = 0.7391$, $Y = 0.2562$, $Z = 0.0047$,
 and $\rm X = 0.7464$, $Y = 0.2515$, $Z = 0.0021$ in order to make
 a comparison with \cite{Brott2011}'s models for the Galaxy, LMC and SMC and make a preliminary test of our results
 respectively. All other elements (including C, N, O, Mg, Si, Fe)
 follow the solar abundances in \cite{Asplund2005}.

\section{Results}
Based on \cite{Meynet2003}, whether a single star
can evolve into a WR star, its rotating velocity
is crucial. Therefore, we take different initial
rotating velocities to investigate its effects on stellar
evolution. In order to compare with results in \cite{Brott2011},
we set the specific initial surface velocities $v_{\rm i}=0$, 250, 500 and 650 km ${\rm s^{-1}}$
in different models, respectively.

\subsection{Evolutionary tracks}
\begin{figure*}
\begin{tabular}{lr}
\subfigure[] {\includegraphics[height=3.5in,width=3.5in,angle=-90]{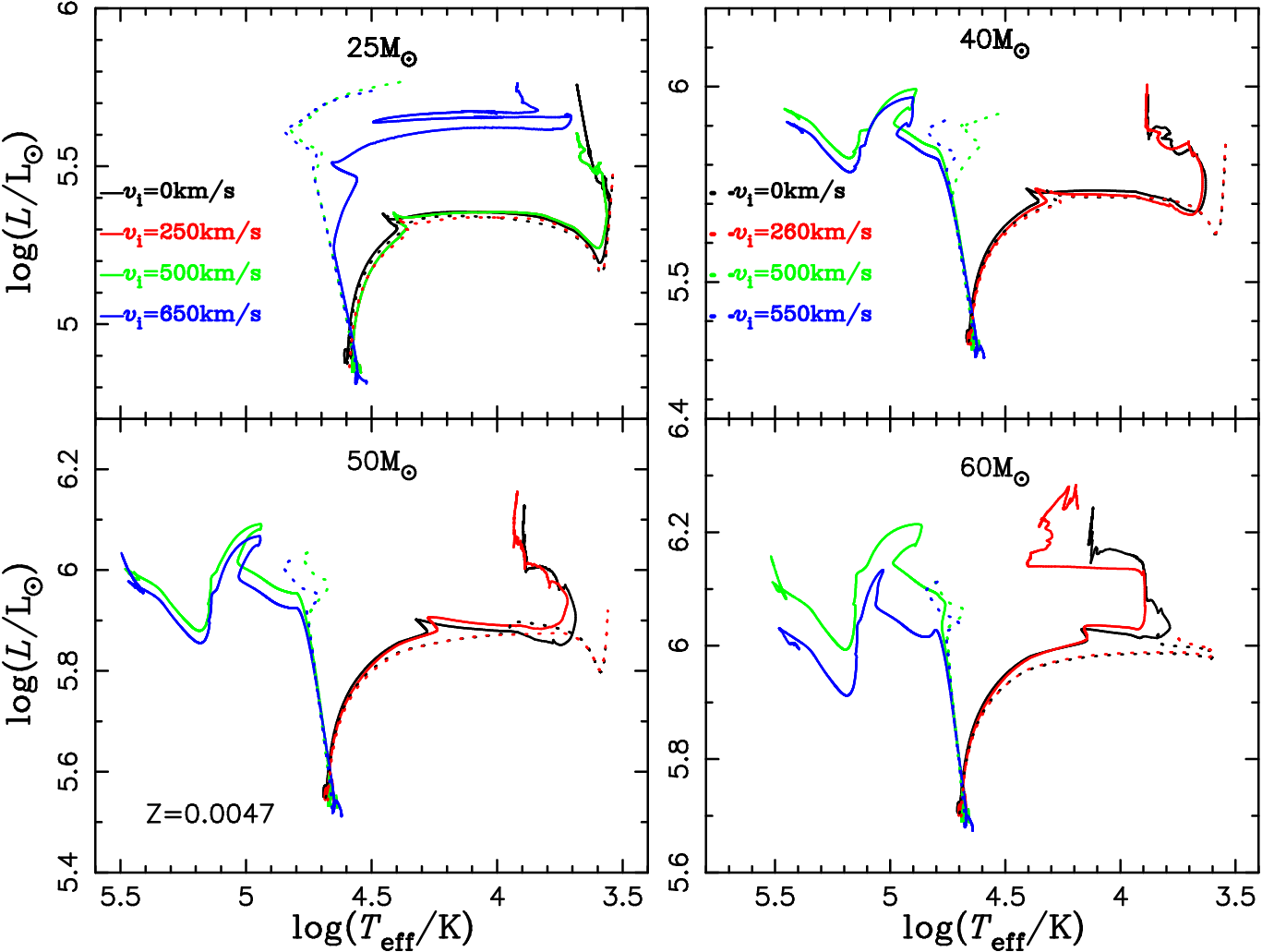}}&
\subfigure[] {\includegraphics[height=3.5in,width=3.5in,angle=-90]{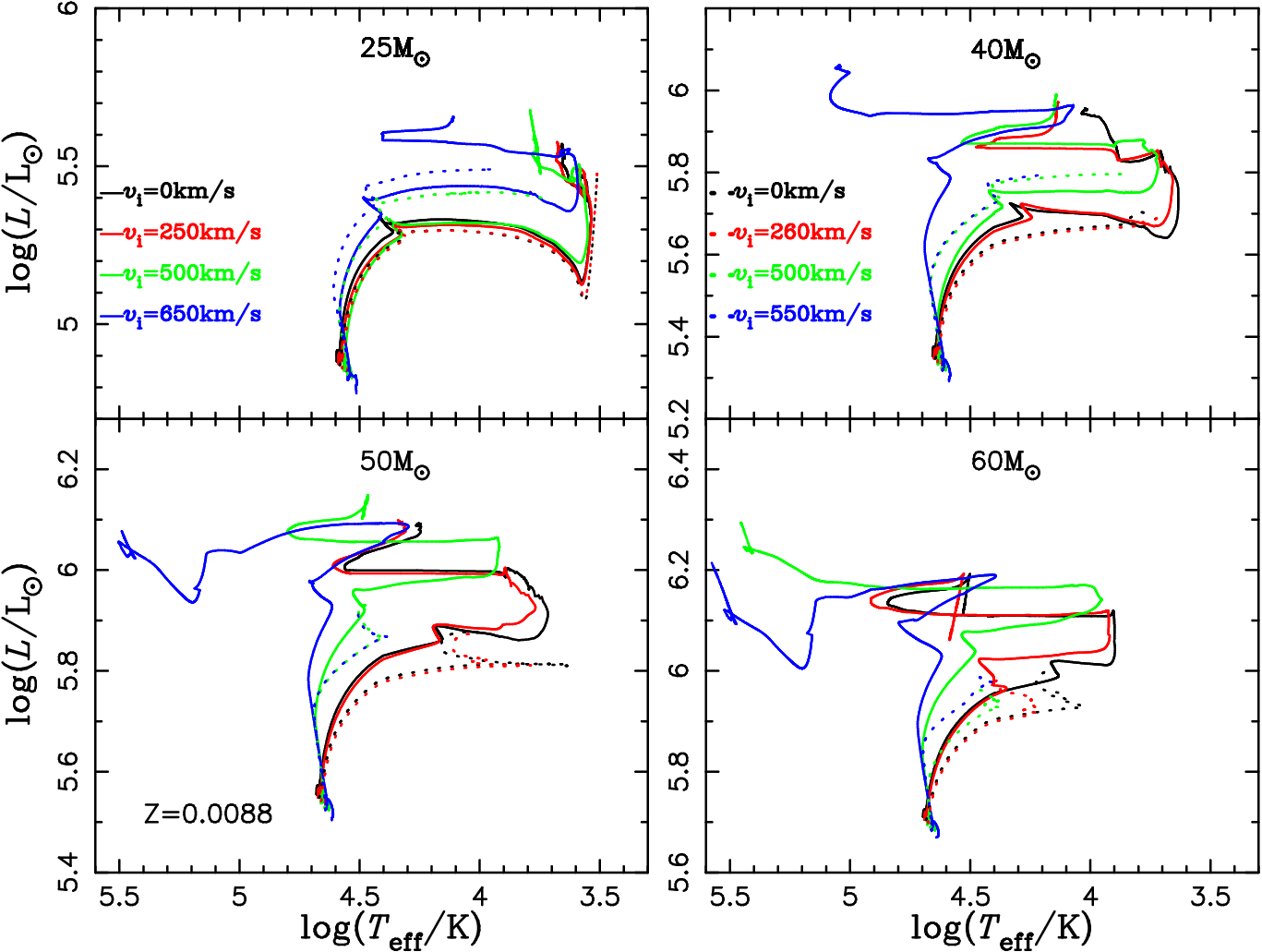}}\\
\end{tabular}
\caption{Evolutionary tracks compared with \cite{Brott2011} for Z=0.0047 (left) and Z=0.0088 (right) for mass of 25, 40, 50, 60 $M_{\odot}$ from left-top to right-bottom . The solid colorful lines are our models and the dotted colorful lines represent the models of \cite{Brott2011}, different $v_{\rm i}$ is represented by different colors which are shown in the legend. }
\label{fig1}
\end{figure*}

Figure \ref{fig1} shows the evolutionary tracks in HR diagram compared with \cite{Brott2011}
 with the initial masses of 25, 40, 50 and 60 $M_\odot$, the
initial rotating velocities of 0, 250, 500 and 650 km ${\rm s^{-1}}$
and the metallicities of 0.0047 and 0.0088, respectively.
Obviously, the rotating velocity have great effects on the evolution of the
massive stars. For rapid rotating models,
owing to the mixing timescale becomes shorter than nuclear timescale,
the materials produced in the core are
transported to the outer layers and mixed, resulting in (quasi-)
chemically homogeneous \citep{Maeder1987, Langer1992,Martins2013}.
At the same time, the large amount of He in the outer layers reduces the opacity,
causing the star hotter, hence, the stars which evolved chemically homogeneously
show a blue tendency in the HR diagram from the main sequence, just as shown in Figure \ref{fig1}
for models with $v_{\rm i}$ = 500 and 650 km ${\rm s^{-1}}$.
Meanwhile, the chemically homogeneous caused by rotation also depend on the metallicity.
Compared the left panel with the right panel in Figure \ref{fig1},
the smaller the metallicity is, the stronger the blue tendency is.

\cite{Brott2011} only calculated the evolution of the rotating massive stars
on main sequence. In this work, we compute the evolution from the beginning of hydrogen burning to the
end of carbon and oxygen burning and enlarge the range of the metallicity.
Considering, the input parameters in this paper are similar to \cite{Brott2011},
we compare our evolutionary tracks with these in \cite{Brott2011}.
There are some differences, especially for stars at higher mass and rotating velocity.
They may result from some uncertainties on
simulating massive stars with high rotating velocity in the different codes \citep{Zhu2017}. For instance, the different opacity used between us and \cite{Brott2011}, the lower opacity in our models contributes to the fact that they are
hotter and more compact \citep{Cantiello2009,Gotberg2017}.

\subsection{Evolve to WR stars}
\begin{figure*}
\begin{tabular}{lr}
\includegraphics[totalheight=3.5in,width=3.2in,angle=-90]{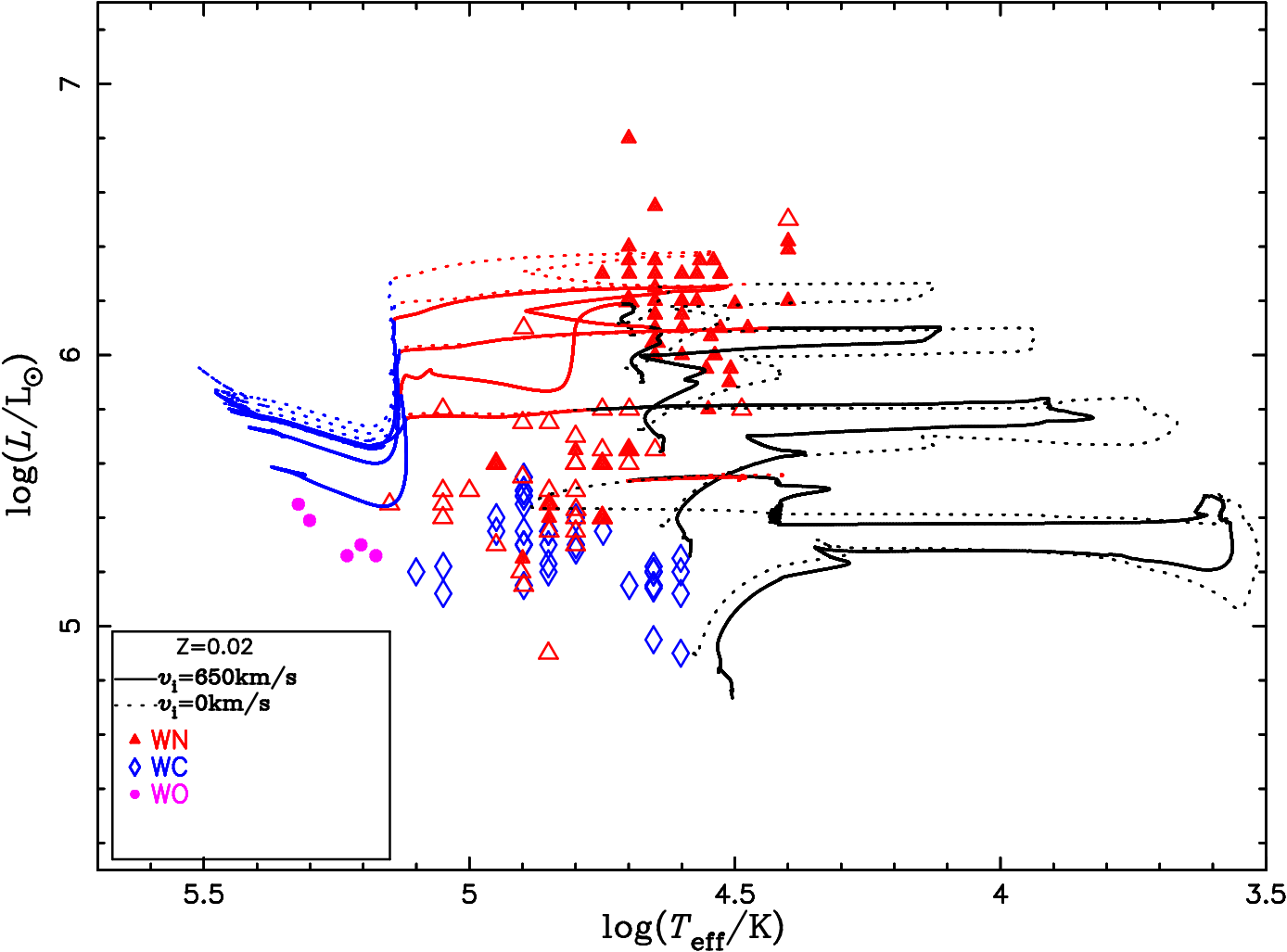}&
\includegraphics[totalheight=3.5in,width=3.2in,angle=-90]{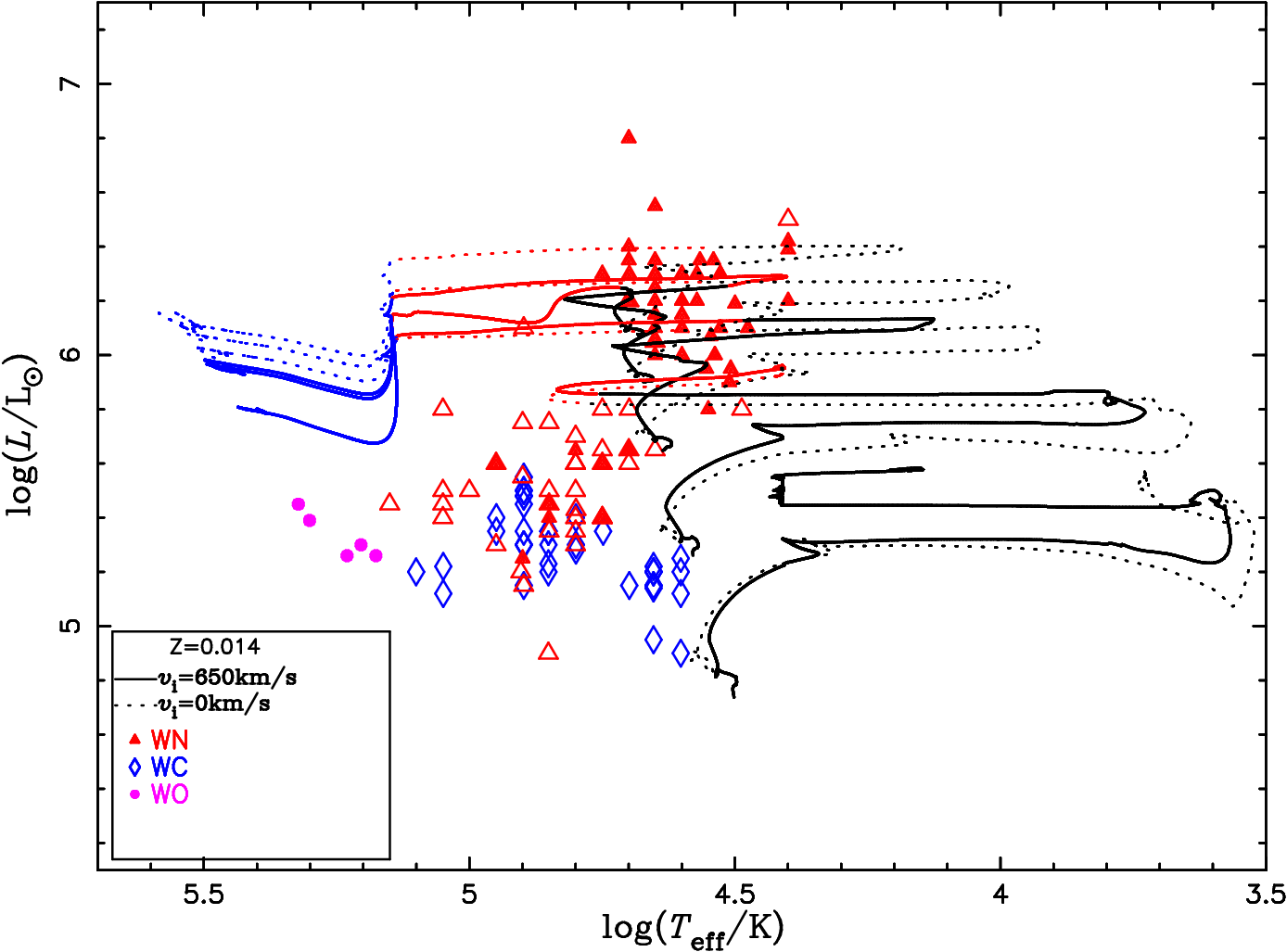}\\
\end{tabular}
\caption{The positions of WRs observed in the MW and the evolutionary tracks of
massive stars with different masses (from the bottom to the top the stellar masses are
25, 40, 60, 80 and 100 $M_{\rm \odot}$, respectively).
The black lines represent the pre-WR phase ( defined as $X_{\rm s}>0.3$),
the red lines for WN ($10^{-5}<X_{\rm s}<0.3$), while the blue lines for WC sequence ($X_{\rm s}<10^{-5}$).
The dashed and solid lines represent non-rotating and rapidly rotating models, respectively.
Different types of WR stars are showed with different icons given in the legend.
The observational data of WN, WC and WO stars in the Galaxy come from \cite{Hamann1995},
  \cite{Hamann2006a}, \cite{Liermann2010}, \cite{Sander2012}, \cite{Martins2008} and \cite{Tramper2015}, respectively.}
\label{mw}
\end{figure*}

\begin{figure*}
\begin{tabular}{lr}
\includegraphics[totalheight=3.5in,width=3.2in,angle=-90]{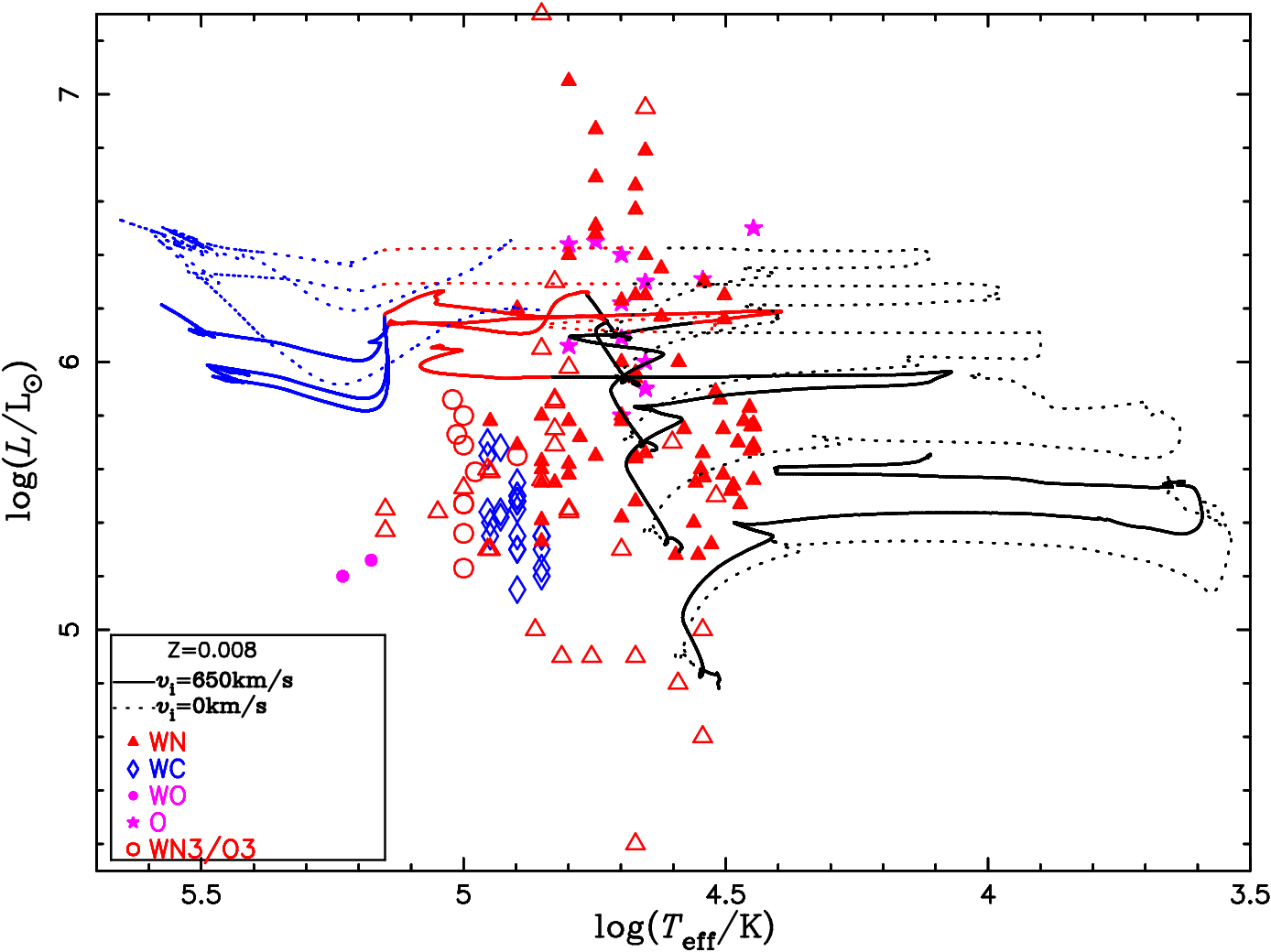}&
\includegraphics[totalheight=3.5in,width=3.2in,angle=-90]{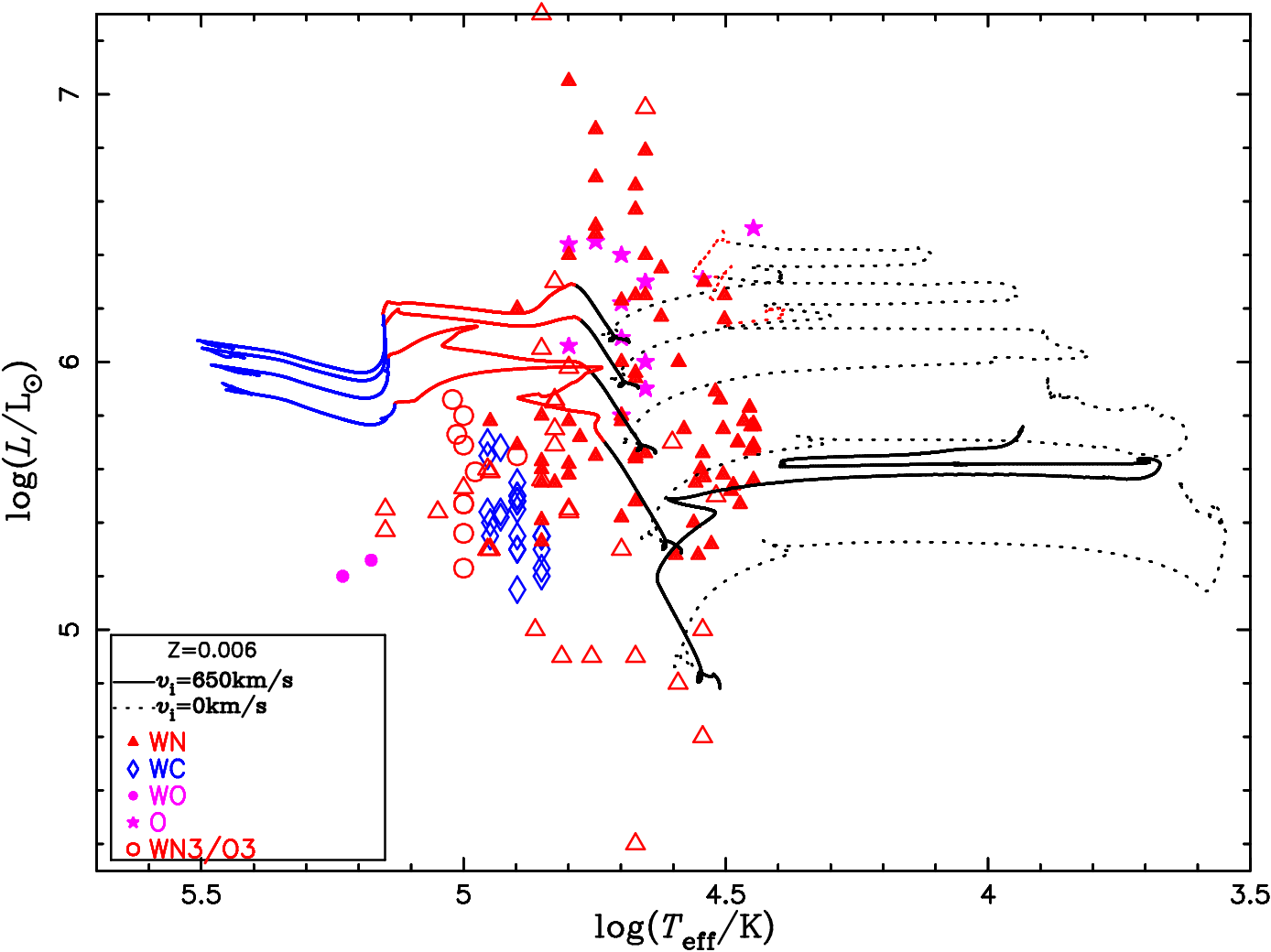}\\
\end{tabular}
\caption{The same as Figure \ref{mw} but for Z=0.008(left) and 0.006(right). The observational data of WC and WN stars in the LMC originate from \cite{Crowther2002},
\cite{Hainich2014}, \cite{Crowther1997}, \cite{Tramper2015} and \cite{Koesterke1991}.
The nine WN3/O3s observed by \cite{Neugent2017} are also included.
}\label{lmc}
\end{figure*}

\begin{figure*}
\begin{tabular}{lr}
\includegraphics[totalheight=3.5in,width=3.2in,angle=-90]{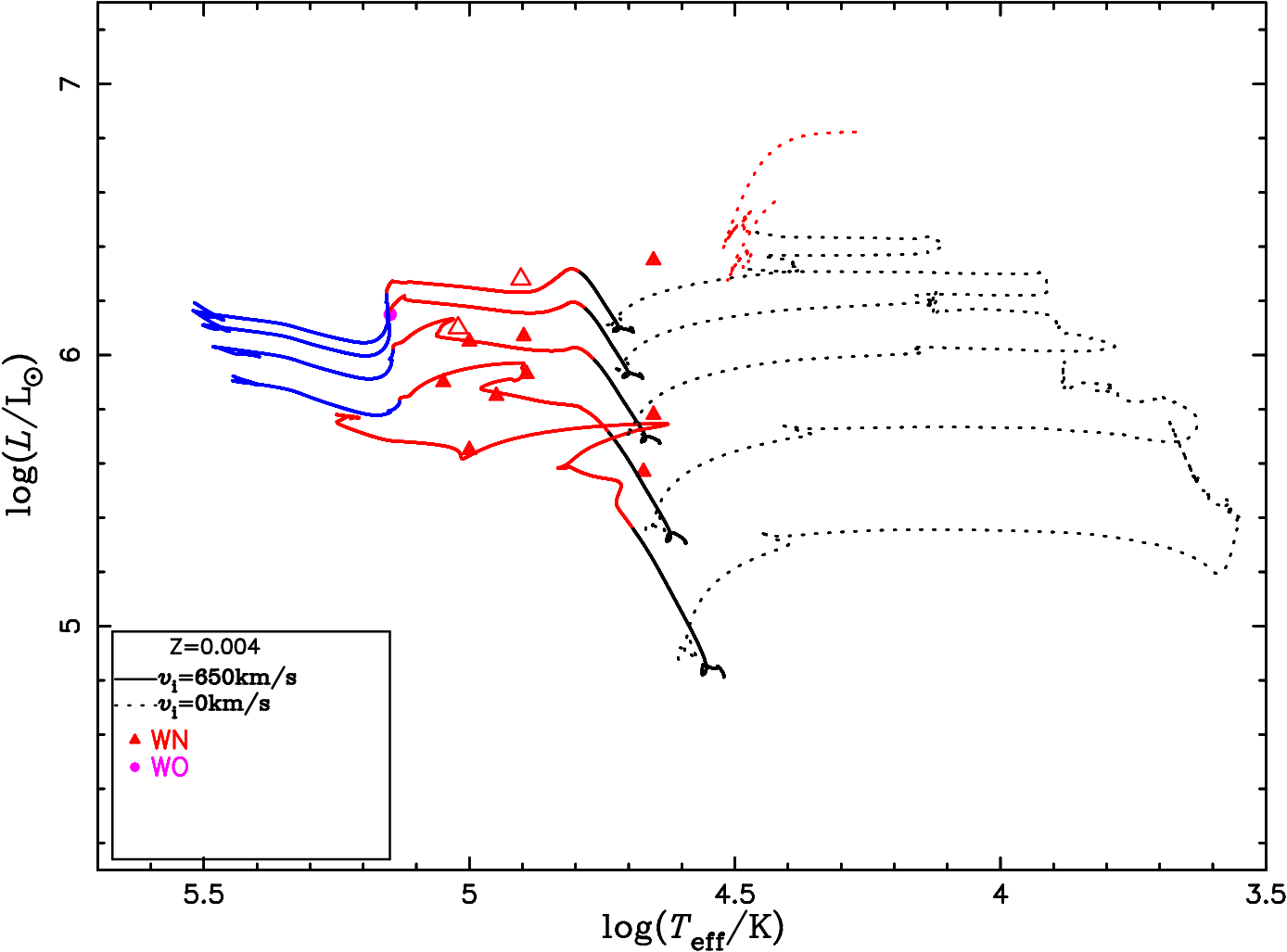}&
\includegraphics[totalheight=3.5in,width=3.2in,angle=-90]{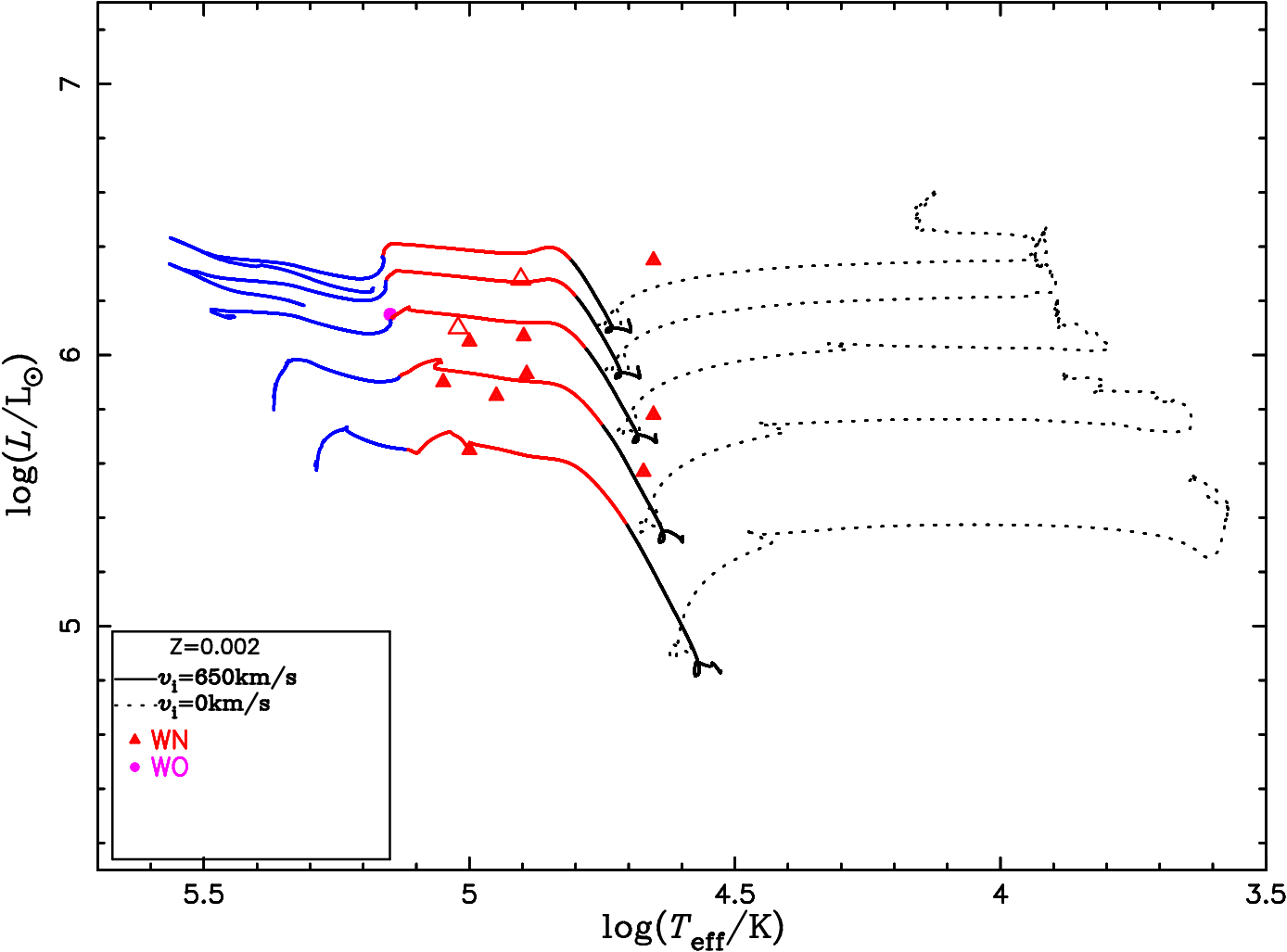}\\
\end{tabular}
\caption{The same as Figure \ref{mw} but for Z=0.004(left) and Z=0.002(right). And for WR stars in SMC, the parameters adopted here are from \cite{Hainich2015} and \cite{Shenar2016}.
}\label{smc}
\end{figure*}

Most of WR stars observed are in the Galaxy and the LMC, only twelve are from SMC.
They are plotted in Figure \ref{mw}, Figure \ref{lmc} and Figure \ref{smc}.
The observational data of WN, WC and WO stars in the Galaxy come from \cite{Hamann1995},
  \cite{Hamann2006a}, \cite{Liermann2010}, \cite{Sander2012}, \cite{Martins2008} and \cite{Tramper2015}.
The observational data of WC and WN stars in the LMC originate from \cite{Crowther2002},
\cite{Hainich2014}, \cite{Crowther1997}, \cite{Tramper2015} and \cite{Koesterke1991}.
The nine new type of WR stars WN3/O3 stars in the LMC observed by \cite{Neugent2017} are also included,
which spectroscopically resemble a WN3 and O3V binary systems but visually too faint to be WN3+O3V binary systems.
Despite some of the stars listed with a WN classification in the BAT99
catalog have been reclassified as O-types stars\citep[e.g.,][]{Taylor2011,Niemela2001,Crowther2011,Evans2011},
they are also collected in our sample, and we call them as O-type stars in the following sections.
In the SMC, almost all known WR stars belong to the WN sequence excepting the binary
system SMC AB 8 whose primary belongs to WO type, and their parameters adopted here
are from \cite{Hainich2015} and \cite{Shenar2016}.
The hydrogen-rich and hydrogen-free WN stars are separated out
with solid triangle and hollow triangle in HR diagrams, respectively.

Considering the distribution of
 metal abundance in galaxies is inhomogeneous, for instance, the metallicities in
 galactic disk are higher than that in galactic nucleus and halo, and the
 strongly dependence of metallicities on galactic age, for example,
 the young massive stars in the LMC may reach nearly solar values \citep{Piatti2013,Hainich2014}.
 We adopt the two thresholds of metallicities for each galaxies.
 Figure \ref{mw}, Figure \ref{lmc} and Figure \ref{smc}
show the evolutionary tracks of non-rotating and
rapidly rotating massive stars for
different $Z$s.

Based on the works of \cite{Smith1991}, \cite{Meynet2003} and \cite{Groh2013},
a massive star evolves into a WR star when the hydrogen abundance ($X_{\rm s}$) around its surface is less than 0.3,
and it may become a late-type WR star (WC star($X_{C}>X_{N}$ , and surface abundances
(by number) such as $\frac{C+O}{He}<1$) or WO star($X_{C}>X_{N}$ and $\frac{C+O}{He}<1)$)) when $X_{\rm s}<10^{-5}$.
According to Figure \ref{mw}, WN stars with low effective temperature
can originate from massive stars at high metallicity.
For these massive stars,
majority of their angular momentum were taken away by their severe stellar winds,
which can rescue the rotation, and makeing it difficult to produce
efficient chemical homogenous evolution.
There is not significant difference between the evolutionary tracks for
these massive stars evolving into WR stars without rotating velocity and with highly rotating velocity.
However, rapid rotation can lead to efficient
chemical homogenous evolution for the massive stars at low metallicity,
for instance, the LMC and SMC models as shown in Figure \ref{lmc} and \ref{smc}.
The evolutionary tracks of these massive stars without rotation and with rapid rotation are
completely different. The massive stars at low metallicity and low rotating velocity hardly
evolve into WR stars, but these with high rotating velocity rapidly become
WR stars. This can explain the origin of single WR stars in the low-metallicity galaxy.
Their evolutionary tracks pass through the zone covered by WN
stars with high effective temperature in Figure \ref{lmc} and Figure \ref{smc}.
Although these tracks also cover several WC stars, even several WO stars,
our results can hardly explain the origin of the WR stars located in the left-bottom
zone.
However, the binary models computed by \cite{Gotberg2017}
indicating that much of them
may evolved from rapid rotating star that are produced by spin-up
during mass transfer in binary systems.
In addition, we can see that some WN and WC stars are cooler than predicted by our stellar evolution models,
which has been also proposed by \cite{Eldridge2017}, \cite{Hamann2003} and \cite{Sander2012},
this may be attributed to the inflating of the envelope caused by clumping
in the outer convective zone of the star \citep{McClelland2016, Grafener2012}.
Simultaneously, one should notice that
there are the errors for the observations due to
some factors (for example distance:  the distances of the
majority of WR stars are not well established.).

We find that there is a temperature offset between the hydrogen-rich
 WR stars and our quasi-homogenous evolutionary models, this is because
 that hydrogen-rich WR stars usually show signature of hydrogen in their
 atmospheres which is inconsistent with chemically homogeneous
 evolution \citep{Shenar2016}. The nine WN3/O3s discovered by \cite{Neugent2017}
 known as a new class of WR stars are shown in LMC models in
 Figure \ref{lmc} with red circle symbols. We can see that they can be
 mostly but not all reproduced by our rotating models, and our results is similar to that of
 \cite{Eldridge2017} who considered that their both single and binary models
 agree well with the observed luminosity, temperature and surface composition of WN3/O3s.
 This indicates that they may be typical WR stars but with different mass-loss rates.

Besides, as revealed in our paper and the works of \cite{Brott2011} and \cite{Koenigsberger2014},
 rotating stars at low metallicities are more likely to induce chemically homogeneous evolution.
\cite{Koenigsberger2014} proposed that the binary system contained
in multiple system HD 5980 in the SMC is the product of quasi-chemically
homogeneous evolution with little or no mass transfer. However, it is found that
quasi-chemically homogeneous evolution does not seem
consistent with AB 5, either with AB 3, 6 and 7,
since the temperature of the primary is overpredicted by
more than $2\sigma$ \citep{Shenar2016,Eldridge2011,Eldridge2012},
and we can get the same conclusion from Figure \ref{smc}.
This is because of the lower value of $T_{\rm eff}$ adopted in our work
compared to that used by \cite{Koenigsberger2014} \citep{Shenar2016}. \cite{Eldridge2017}
proposed that AB2, whose location matches quite closely to
their $40M_{\odot}$ quasi-chemically homogeneous evolution track, is perhaps the candidate
for a quasi-chemically homogeneous evolution (by mass transfer) star.
Although we $\rm can^{'}t$ get the same conclusion from Figure \ref{smc} directly,
this possibility can not be ruled out cause the velocity (650 km s$^{-1}$) we shown
is too high enough to lead to chemically homogeneous evolution
(compared with 400 km s$^{-1}$ proposed by \cite{Brott2011} and \cite{Heger2000} ).

\section{CONCLUSIONS}
WR stars are very important objects because they are related to the
type Ib/Ic SNe, LGRBs. They also affect the chemical compositions of interstellar
medium. In this work, we investigate the possibility of a single star
evolving into WR star due to rotation and compare our rapid rotation cases with observations in HR diagrams.
The rotation has few effects on the evolution of massive stars at
high metallicity that is because the rotation rate and efficiency of the mixing process is slowed down due to the
enhancements of stellar-winds mass and angular momentum loss,
and these stars become WR stars when the helium in their center is ignited.
However, the mass loss induced spin-down, which stops the
efficient rotational mixing, is reduced at lower metallicity.
Since then rapid rotating massive stars can easily evolve into
WR stars due to the trigger of the rotational induced chemically homogeneous evolution.
From our models, we find that in the SMC the observed WR stars are consistent with
the single-star evolution models. However at higher metallicities our single-star evolution models
can only explain the luminous, hydrogen-rich WN stars and O stars.
In the LMC and the Galaxy all the WC and WO stars are significantly fainter,
and for the WC stars cooler, than our model tracks.
The same is also true for a significant fraction of the WN stars.
It is therefore likely that the majority of these stars are the result
of binary evolution (e.g. \cite{Eldridge2017}).
Perhaps, it may be also because that our models about rapidly rotating
massive stars are still beyond real ones. Simultaneously, the
observational errors (such as distance) also lead to disagreement.
There is a long way to go before we can understand WR stars.

\section*{ACKNOWLEDGEMENTS}
We acknowledge the anonymous referee for careful reading of the paper and constructive criticism.
This work was supported by the National Natural Science Foundation
of China under Nos. 11473024, 11363005, 11763007, 11503008, 11365022, 11703081, 11521303, 11733008 and 11521303£¬
the XinJiang Science Fund for Distinguished Young Scholars under No. QN2016YX0049£¬
and the National Natural Science Foundation of Yunnan Province (No.
2017HC018) and the CAS light of West China Program.

\bibliography{cui}

\label{lastpage}

\end{document}